\documentclass[conference]{IEEEtran}
\IEEEoverridecommandlockouts
\usepackage{cite}
\usepackage{amsmath,amssymb,amsfonts}
\usepackage{algorithmic}
\usepackage{graphicx}
\usepackage{textcomp}
\usepackage{xcolor}
\usepackage{epsfig}
\usepackage{amsfonts}
\usepackage{comment}
\usepackage[linesnumbered,ruled]{algorithm2e}
\usepackage{bm}
\usepackage{float}
\usepackage{enumitem}

\def\BibTeX{{\rm B\kern-.05em{\sc i\kern-.025em b}\kern-.08em
    T\kern-.1667em\lower.7ex\hbox{E}\kern-.125emX}}
\begin{document}

\title{AI-Driven XR Situational Awareness Platform for Urban Crisis Management and Smart Mobility Operations} 

\author{\IEEEauthorblockN{Dimitris Spyridonidis$^{1,\ast}$, Gerasimos Arvanitis$^{1,\ast}$, Konstantinos Moustakas$^{1}$}
\IEEEauthorblockA{\textit{$^{\ast}$AviSense.ai, Patra, Greece} \\\textit{$^{1}$Department of Electrical and Computer Engineering, University of Patras, Patra, Greece} \\ 
arvanitis@avisense.ai,  moustakas@upatras.gr}

}

\maketitle


\section{Introduction} 
\label{sec:intro}
Urban environments are increasingly exposed to complex, dynamic, and interdependent risks, ranging from traffic incidents and infrastructure failures to large-scale crises such as extreme weather events and emergency response scenarios. In such conditions, decision-makers and operators are required to act under time-critical constraints while relying on fragmented, heterogeneous, and often incomplete information \cite{10311086}. The lack of unified situational awareness, combined with limited visibility and disconnected systems, significantly affects response time, coordination efficiency, and overall operational effectiveness \cite{xiang2024v2xreal}. In parallel, modern cities have deployed extensive sensing infrastructures, including surveillance camera networks, Internet of Things (IoT) devices, connected vehicles, and satellite-based observation systems \cite{ZHANG2026104572}. Although these technologies generate vast amounts of data, their exploitation remains limited due to the absence of integrated platforms capable of real-time data fusion, intelligent interpretation, and intuitive visualization \cite{Antoun2025}. As a result, a substantial gap persists between data availability and actionable intelligence, particularly in safety-critical and crisis management applications \cite{11340049}.

To address this challenge, this paper presents an AI-driven XR situational awareness and operational platform designed for urban crisis management and smart mobility operations. The proposed system integrates data from city infrastructure, connected vehicles, and Vulnerable Road Users (VRUs), enabling a comprehensive and real-time understanding of the urban environment. Through a unified operational dashboard and Augmented Reality (AR) interfaces, the platform supports both centralized monitoring and field-level interaction. By enhancing perception, enabling cooperative awareness, and delivering context-aware information, the proposed approach improves decision-making, coordination, and safety across diverse urban scenarios.


\section{Methodology}  
\label{framework}
The proposed system is designed as a scalable and modular platform that integrates heterogeneous data sources into a unified operational environment, supporting both centralized monitoring through a dashboard interface and decentralized access via XR-based visualization tools. This enables seamless interaction between control centers and field users, facilitating coordinated decision-making in complex urban scenarios.

At the data acquisition level, the system leverages multiple sources, including city infrastructure cameras, mobile sensing platforms such as connected vehicles and edge devices, and complementary data streams from satellite observations and positioning systems. Additional contextual information related to traffic conditions, environmental factors, and emergency events is also incorporated. The inclusion of data from VRUs, such as pedestrians and cyclists, ensures a comprehensive representation of the urban environment beyond vehicle-centric approaches.

The collected data is processed through an AI-driven pipeline that performs object detection, tracking, and behavioral analysis. Computer vision and machine learning techniques are used to identify relevant entities and estimate their trajectories under dynamically changing conditions, including variations in lighting, weather, and traffic density. A key feature of the system is the fusion of data from multiple viewpoints, enabling cooperative perception and reducing the impact of occlusions and blind spots.

The processed information is structured into a unified situational model that feeds an operational dashboard, providing real-time monitoring, event detection, and risk assessment through intuitive visualizations. In parallel, XR-based interfaces deliver contextual information directly to field users via AR, enabling real-time perception of hazards and supporting rapid response in critical situations.

\begin{figure*}
	\centering
	\includegraphics[width=0.99\linewidth]{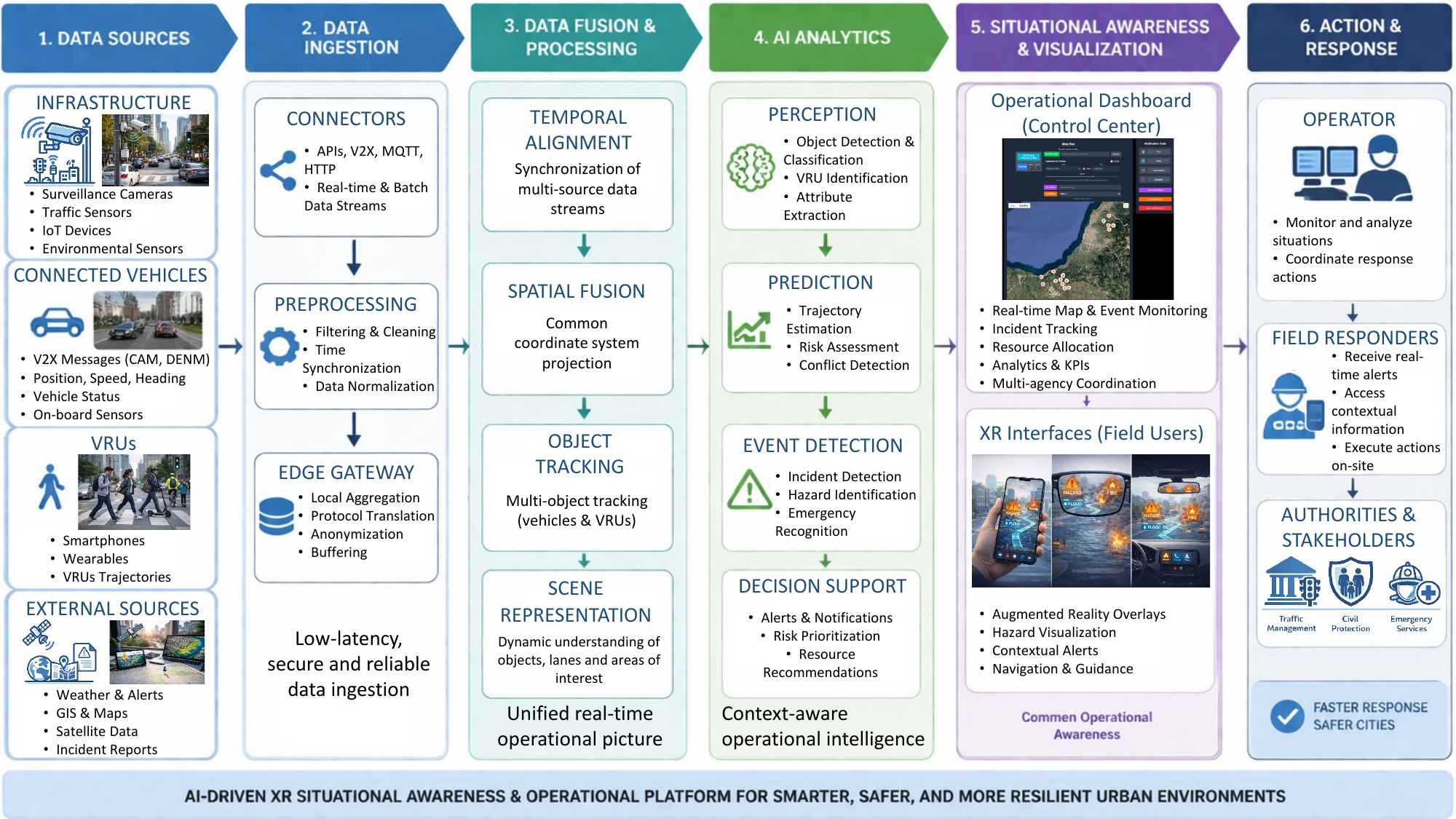}
	\vspace{-0.8em}
	\caption{Architecture of the proposed AI-driven XR situational awareness platform for urban crisis management. The system integrates multi-source urban data, performs real-time fusion and AI-based analysis, and delivers actionable information through operational dashboards and XR interfaces to support coordinated response and decision-making.}
	\label{fig:All}
\end{figure*}

\section{Results and Application Scenarios} 

The proposed platform has been considered within representative urban scenarios, including traffic incident management, pedestrian safety monitoring, and emergency response coordination. The analysis indicates that the system has the potential to significantly enhance situational awareness by integrating heterogeneous data sources and presenting information in a unified and actionable manner.
In traffic-related scenarios, the system enables the detection and continuous tracking of potentially hazardous interactions between vehicles and VRUs. By combining data from different sources, the platform significantly improves spatial coverage (up to 60\%) and reduces the impact of occlusions and blind spots by $\sim$50\%. Detection accuracy for VRUs increases from typical values of 75-80\% to over 90\%, while early hazard detection allows operators to anticipate potential collisions up to 2-3 seconds earlier than baseline systems, leading to a reduction in near-miss incidents by an estimated 25-30\%.
In crisis management scenarios, like emergency response operations, the platform supports real-time monitoring of affected areas and facilitates coordination among multiple stakeholders. The operational dashboard reduces incident identification and assessment time by $\sim$30-40\%, while improving resource allocation efficiency (up to 25\%). At the same time, XR-based visualization tools provide field responders with contextual, real-time information, reducing navigation errors and decision latency by $\sim$20-30\% in complex environments.
Performance evaluation further indicates improvements in system responsiveness, with end-to-end latency maintained $<$100 ms. The integration of real-time data fusion and intuitive visualization mechanisms is expected to contribute to identifying risks earlier, responding more effectively, and maintaining a higher level of operational understanding compared to traditional monitoring approaches.

\section{Discussion and Conclusions} 

The integration of AI-driven data processing, cooperative perception, and XR-based visualization introduces a novel paradigm for urban crisis management and mobility operations. The proposed platform effectively bridges the gap between large-scale data collection and actionable intelligence by leveraging existing city infrastructures and augmenting them with advanced analytics and visualization capabilities. This transition is critical for enabling real-time understanding and response in complex and dynamic urban environments.
In contrast to traditional approaches that rely on isolated systems and fragmented data sources, the presented solution adopts a holistic perspective, integrating information from infrastructure sensors, connected vehicles, and VRUs into a unified operational framework. This comprehensive fusion of heterogeneous data enhances the accuracy and completeness of situational awareness, allowing operators to better interpret ongoing events and coordinate actions across multiple stakeholders. The proposed approach highlights the potential to improve safety and operational efficiency by enabling earlier identification of risks, supporting faster response processes, and facilitating more informed decision-making. The combination of centralized dashboard-based monitoring and decentralized XR-based interaction further strengthens the system’s ability to support both strategic and on-field operations.
Nevertheless, several challenges remain for large-scale deployment, including system scalability, real-time processing constraints, data privacy considerations, and interoperability with existing urban infrastructures and standards. Addressing these challenges is essential for ensuring robustness and adoption in real-world settings.
Future work will focus on enhancing predictive analytics capabilities, integrating next-generation communication technologies (5G/6G), and validating the platform through large-scale pilot deployments, contributing significantly to the development of resilient, safe, and intelligent urban ecosystems.

\section*{Acknowledgement}
The XRay Mobility has indirectly received funding from the Smart Networks and Services Joint Undertaking (SNS JU) under the European Union's Horizon Europe research and innovation programme, via the 6G-PATH –Open Call $\#2$ issued and executed under the 6G-PATH project (Grant Agreement no. 101139172).

\bibliographystyle{IEEEbib}
\bibliography{refs}

@ARTICLE{10311086,
	author={Arvanitis, Gerasimos and Stagakis, Nikolaos and Zacharaki, Evangelia I. and Moustakas, Konstantinos},
	journal={IEEE Transactions on Intelligent Transportation Systems}, 
	title={Cooperative Saliency-Based Pothole Detection and AR Rendering for Increased Situational Awareness}, 
	year={2024},
	volume={25},
	number={5},
	pages={3588-3604},
	doi={10.1109/TITS.2023.3327494}}

@Article{Antoun2025,
	author={Antoun, Maya
		and Elhajj, Imad H.
		and Sayour, Malak
		and Papadopoulos, Panos K.
		and Zarpalas, Dimitrios
		and Rizk, Anthony
		and Islam, ABM Tariqul
		and Huerta, Ivan
		and Toledo, Leonel
		and Arvanitis, Gerasimos
		and Moustakas, Konstantinos
		and Jevti{\'{c}}, Aleksandar
		and Agust{\'i}, Isaac
		and Vona, Francesco
		and Voigt-Antons, Jan-Niklas
		and Warsinke, Maximilian
		and Bailer, Werner
		and Thallinger, Georg
		and Howkins, Ben
		and Poder, Irma
		and Ruess, Maj
		and Asmar, Daniel},
	title={Interactive digital twins enabling responsible extended reality applications},
	journal={Scientific Reports},
	year={2025},
	month={Oct},
	day={03},
	volume={15},
	number={1},
	pages={34539},
	issn={2045-2322},
	doi={10.1038/s41598-025-17855-9},
	url={https://doi.org/10.1038/s41598-025-17855-9}
}

@article{xiang2024v2xreal,
	title={V2X-Real: A Large-Scale Dataset for Vehicle-to-Everything Cooperative Perception},
	author={Xiang, Hao and Zheng, Zhaoliang and Xia, Xin and others},
	journal={arXiv preprint arXiv:2403.16034},
	year={2024}
}

@article{ZHANG2026104572,
	title = {What2Keep: A communication-efficient collaborative perception framework for 3D detection via keeping valuable information},
	journal = {Computer Vision and Image Understanding},
	volume = {263},
	pages = {104572},
	year = {2026},
	issn = {1077-3142},
	doi = {https://doi.org/10.1016/j.cviu.2025.104572},
	url = {https://www.sciencedirect.com/science/article/pii/S1077314225002954},
	author = {Hongkun Zhang and Yan Wu and Zhengbin Zhang}
	}

@INPROCEEDINGS{11340049,
  author={Krasniqi, Edona and Shehu, Visar},
  booktitle={2025 IEEE International Conference on Metrology for eXtended Reality, Artificial Intelligence and Neural Engineering (MetroXRAINE)}, 
  title={A Survey on Extended Reality in Disaster Management: Integrating XR, IoT, and Machine Learning for Enhanced Situational Awareness}, 
  year={2025},
  volume={},
  number={},
  pages={547-552},
  doi={10.1109/MetroXRAINE66377.2025.11340049}}

\end{document}